# COMPUTATIONAL MODEL COMBINED WITH IN VITRO EXPERIMENTS TO ANALYSE MECHANOTRANSDUCTION DURING MESENCHYMAL STEM CELL ADHESION

Jean-Louis Milan[1], Sandrine Lavenus[2,4,*], Paul Pilet[3], Guy Louarn[4], Sylvie Wendling[1], Dominique Heymann[2], Pierre Layrolle[2] and Patrick Chabrand[1]

[1]Institute of Movement Sciences (ISM), Aix-Marseille Univ, CNRS UMR 7287, F-13288 Marseille, France
[2]Inserm, U957, Laboratory of Physiopathology of Bone Resorption (LPRO), Faculty of Medicine, University of Nantes, France
[3]Laboratory of Osteoarticular and Dental Tissue Engineering (LIOAD), Inserm U791, Faculty of Dental Surgery, University of Nantes, France
[4]Institut des Matériaux Jean Rouxel (IMN), CNRS UMR6502, University of Nantes, France

## Abstract

The shape that stem cells reach at the end of adhesion process influences their differentiation. Rearrangement of cytoskeleton and modification of intracellular tension may activate mechanotransduction pathways controlling cell commitment. In the present study, the mechanical signals involved in cell adhesion were computed in *in vitro* stem cells of different shapes using a single cell model, the so-called Cytoskeleton Divided Medium (CDM) model. In the CDM model, the filamentous cytoskeleton and nucleoskeleton networks were represented as a mechanical system of multiple tensile and compressive interactions between the nodes of a divided medium. The results showed that intracellular tonus, focal adhesion forces as well as nuclear deformation increased with cell spreading. The cell model was also implemented to simulate the adhesion process of a cell that spreads on protein-coated substrate by emitting filopodia and creating new distant focal adhesion points. As a result, the cell model predicted cytoskeleton reorganisation and reinforcement during cell spreading. The present model quantitatively computed the evolution of certain elements of mechanotransduction and may be a powerful tool for understanding cell mechanobiology and designing biomaterials with specific surface properties to control cell adhesion and differentiation.

*Address for correspondence:
Sandrine Lavenus
Rensselaer Polytechnic Institute,
Nanotechnology Center and CBIS,
110 Eighth Street, MRC-210,
Troy, NY 12180-3590, USA
        Telephone Number: (518) 276-4217
        FAX Number: (518) 276-6540
        E-mail: lavens@rpi.edu

## Introduction

Cell adhesion is fundamental in the field of biomaterials. Especially in orthopaedics, it is crucial for prosthesis osseointegration (Lavenus *et al.*, 2010). Osteoblasts need to adhere on the implant to synthesise the bone tissue that is directly connected to it without the intervening soft tissue. Moreover, stem cells, which are recruited to heal the implant interface, adhere on the surface and differentiate depending, among other factors, on their adherent morphology. For instance, stem cells which reach at the end of adhesion round, elongated or branched shapes depending on substrate topography or stiffness, differentiate preferentially into adipocytes, fibroblasts or osteoblasts, respectively (McBeath *et al.*, 2004; Engler *et al.*, 2006). Therefore, biomaterial surface must promote adhesion of stem cells and their differentiation into the desired phenotype so as to synthesise a periprosthetic tissue, ensuring efficient and viable biomaterial integration.

Biomaterials that mimic the natural composition of the extracellular matrix have been developed. They are composed of a biological phase or are coated with RGD complex and showed positive effects on cell adhesion (Von der Mark *et al.*, 2010). Topography and surface chemistry, which play a key role in cell adhesion, may be controlled using nanotechnology. This announces the development of new implant surfaces with predictable tissue-integrative properties (Le Guéhennec *et al.*, 2007; Lavenus *et al.*, 2010).

Cell adhesion results from the creation of several discrete focal adhesion points (FAPs). FAPs are each composed of several physical and chemical interactions between proteins or non-biological molecules of the substrate and cellular transmembrane complexes such as integrins (Van der Flier and Sonnenberg, 2001). FAPs bind the cytoskeleton (CSK) mechanically to the biomaterial substrate. FAPs have high tensile strength and are able to withstand the CSK actin-myosin contraction that cells produce to increase their stiffness and stability. The cell adhesion process comes about through an extension of actin protrusions, so-called lamellipodia and filopodia that try to establish remote FAPs. Then, acto-myosin filaments assemble into stress fibres, which interconnect FAPs and anchor and pull the cell on the substrate (Murphy-Ullrich, 2001). Cell adhesion involves mechanotransduction. Changes in cell shape alter Rho

A-mediated CSK contractility, focal adhesion assembly and downstream integrin signalling (Itano *et al.*, 2003; McBeath *et al.*, 2004; Wang *et al.*, 2009; Kilian *et al.*, 2010). These processes required a tensile CSK. FAPS, as tension-dependent structures, need tensile force from CSK to maintain and mature. In addition, numerous studies have shown a determinant role of CSK tonus on cell fate (McBeath *et al.*, 2004; Bhadriraju *et al.*, 2007; Cohen *et al.*, 2007; Kilian *et al.*, 2010). Itano *et al.* reported that tensile forces transmitted by the CSK to the nucleus during adhesion led to nuclear membrane deformation, ion channel opening and then entry of calcium, which induced the transcription of specific genes (Itano *et al.*, 2003). Understanding the cell differentiation that follows cell adhesion requires identifying the mechanotransduction pathways *via* the CSK and nucleoskeleton (NSK) in taking into account the role of intracellular tonus that appears as a signal activating mechanosensitive complexes.

Several authors have studied experimentally the mechanics of cell adhesion by determining the magnitude of force at FAPs and the distribution of stress fibres (Green *et al.*, 1986; Jean, 1999; Deguchi *et al.*, 2006; Milan *et al.*, 2007). Mechanical models based on tensegrity have also been proposed to relate the internal tension to the adhesion conditions (Kurachi *et al.*, 1995; Balaban *et al.*, 2001; Riveline *et al.*, 2001; Brangwynne *et al.*, 2006; Dubois and Jean, 2006; McNamara *et al.*, 2010). These models predict deformation and spatial rearrangement of CSK filaments. However, these models could not describe completely the reorganisation of the CSK involving (dis)assembling of filaments due to fixed connectivity between constitutive elements. In a previous study, we combined the tensegrity concept with divided medium mechanics in order to develop the Cytoskeleton Divided Medium (CDM) model (Del Rio *et al.*, 2009). This CDM model represents the pre-stressed structure of the CSK with an equilibrated system of multiple tensile and compressive interactions. During CDM model deformation, interactions can appear or vanish according to their mechanical properties, which introduce changes in model connectivity and predict CSK deep reorganisation. The various CSK filament networks, such as stress fibres, sub-membrane actin cortex, microtubules and intermediate filaments were included in the model by using a biomimetic approach and their contribution to the overall mechanical behaviour of the cell was analysed.

Since adherent mesenchymal stem cells may differentiate following the shape they reach, we proposed here to compute and identify the mechanical signals inside *in vitro*-cultured stem cells of different adherent shapes. Thus, we developed a three-dimensional CDM model that includes a deformable nucleus with NSK to analyse in each adherent cell the level of intracellular tonus and its effect on the nuclear deformation.

The CDM model was also implemented to simulate adhesion and spreading of a round cell in suspension that entered into contact with a flat substrate coated with adhesion proteins. Cell spreading was simulated by emission of filopodia and creation of remote FAPs. The morphology and mechanical state of the cell model obtained *via* this so-called free adhesion simulation were then compared to those of cultured cells. We hypothesised that the intracellular tonus of adherent cells depends on cell morphology and induces a level of nuclear deformation that in case of stem cells may lead to a specific phenotype.

**Materials and Methods**

**Experimentation**

*Cell culture*
Human mesenchymal stem cells (hMSC) were obtained from bone marrow cell aspirations harvested from the iliac crest of patients undergoing orthopaedic surgery and after receiving informed consent (Etablissement Francais du Sang, Centre-Atlantique, Tours, France). Bone marrow cells were plated in 75 cm$^2$ tissue culture flasks (Corning) with 15 mL of culture medium. Standard culture medium consisted of alpha modified Eagle medium (α-MEM, Gibco, Invitrogen, Carlsbad, CA, USA) supplemented with 10 % foetal calf serum (FCS, Biotech, Aidenbach, Germany), 2 mM of L-glutamine, 100 µg/mL of streptomycin and 100 Units/mL of penicillin. The culture medium was refreshed every 2 days and adherent hMSCs were cultured until 80 % confluence in a humidified atmosphere, 5 % $CO_2$ at 37 ºC. Cells were detached with trypsin/EDTA (0.05 % v/v, Gibco, Invitrogen) for 5 min and counted on a Malassez's haemocytometer by using trypan blue exclusion dye.

*Cell adhesion*
Cells were seeded on tissue culture polystyrene (TCPS) into 24-well plates (Corning, Invitrogen) at a densities of 10,000 cells/cm$^2$. Three samples were used and the experiments were carried out at least three times. Cells were counted and diluted in culture medium without osteogenic factor. Cell suspensions (500 µL) were rapidly (less than 10 min) poured into each well. Cells were allowed to adhere to the substrate for 4 h at 37 ºC in humidified air with 5 % $CO_2$.

*Actin, vinculin and nuclear staining*
After removing the culture medium, the cells were rinsed three times with phosphate buffered saline (PBS), fixed with 4 % paraformaldehyde (Sigma, St. Quentin Fallavier, France) for 20 min and then washed three more times to remove excess paraformaldehyde. Samples were stored at 4 ºC until staining for actin, vinculin and nuclei. Fixed cells were first permeabilised with 0.5 % Triton (x100, Sigma) in PBS at 4 ºC for 15 min. In order to reduce non-specific background, samples were blocked with PBS/bovine serum albumin (BSA) 1 % (Sigma) for 10 min at 37 ºC. After blocking, the PBS/BSA was aspirated and the samples were first incubated for 1 h with Alexafluor 488-phalloidin (Molecular Probe, Invitrogen) at a dilution of 1:40 in 10 mg/mL BSA in PBS at 37 ºC in the dark. After each incubation, the samples were rinsed twice with 0.05 % PBS/Tween (Sigma). Anti-vinculin (Monoclonal anti-vinculin clone hVIN-1 produced in mouse, V9131, Sigma) at a dilution of 1:100 in PBS/BSA 1 % was added to each well for 1 h at 37 ºC in the dark and 30 min with secondary rabbit anti-mouse antibody (A11061,

Molecular Probe, Invitrogen) at a dilution of 1:200 at room temperature. Finally, samples were incubated with Hoechst (Sigma) for 10 min at room temperature, and washed twice with deionised water. Samples were stored at 4 ºC before fluorescence imaging (Nikon Eclipse, TE2000E, Champigny Sur Marne, France).

*Image analysis of cell morphology*
After staining the actin, vinculin and nuclei, fluorescent images were analysed using a custom-made program with image analysis system (Quantimet Q550, Leica Microsystems, Wetzlar, Germany) as previously described (Lavenus *et al.*, 2011). Four images were recorded at a magnification of x10 and x20 for each staining. A semi-automatic binary treatment was performed on each image. The number of cells attached to the substrates was automatically counted using nuclear staining. Actin staining was used for cataloguing cell shapes into round, elongated or branched morphologies. For insight into the cells spreading on surfaces, cell areas and the number of FAPs were analysed. For each morphology category, a range of 80 to 300 cells was analysed among which one specimen of cell was chosen for the computation.

**Computational model**
*Constitutive equations*
The present model was a three-dimensional extension of a previous one, the 2D CDM model (Milan *et al.*, 2007) in benefiting some improvements. It was based on the mechanics of multi-interaction system derived from theory of divided or granular mediums, i.e., a set of particles in contact. Contrary to classical finite element models that describe the cell as a continuum and do not represent CSK filamentous structure, the CDM model describes the cell as a set of particles that interact with each other and generate a discrete force network able to mimic in cell the discrete filament network of CSK.

The CDM model represents in its initial state a 15 μm-diameter round cell with a 6 μm-diameter nucleus. Cell volume was divided automatically into a set of particles using a custom-made pre-processor. The set was generated as polydisperse, i.e., composed of particles of various sizes so as to avoid crystal ordering and to facilitate particle movement. Particle centres defined the nodes of the divided medium and of the network of interactions. The average free distance between two CSK nodes was estimated and set in the CDM model at d = 0.5 μm. This resulted in fixing the average particle diameter at d, with diameters in the range of 0.4-0.8 μm. As a result, the whole cell volume was modelled by a compact and polydisperse set of roughly 12,000 rigid spherical particles composing the divided medium (Fig. 1). In the middle of the cell, 330 spheres were aggregated and formed the nuclear core (Fig. 1a). The nuclear core was enclosed in a layer of 340 spheres that formed the nuclear membrane. 50 perinuclear spheres were placed following a regular pattern all around the nucleus at a distance of 2 μm from the nuclear membrane (Fig. 1b). 8,300 spheres were aggregated around the nucleus and formed the cell core composing the main part of the divided medium (Fig. 1c). Finally, the cell core was surrounded by a layer of 2,600 spheres that formed the cell membrane (Fig.

1d). Fig. 1e shows the CDM model at the beginning of free adhesion simulation on a flat substrate. At that point, the model had 150 potential adhesion complexes located in the membrane and able to connect extracellular proteins distributed randomly on the substrate with a density of 3.5 proteins/μm$^2$ (Fig. 1e).

The nodes of the medium were classified into species: nuclear core, nuclear lamina, perinuclear, cell core, cell membrane, adhesion complexes and extracellular proteins, between which specific interaction laws were defined. Interaction laws were written as a relation between reaction force and the gap between nodes. Two interaction laws, known as Elastic Wire and Contact and described in our previous study, were used here (Milan *et al.*, 2007). Elastic Wire law acted like a virtual pre-strained elastic wire between two nodes: tension was proportional to stretching and became null when the elastic wire slackened. As shown in the set of Eq. 1, the tensile reaction force $R$ between two nodes was given as a function of the gap $g$, $g_0$ being the gap at the beginning of the simulation, $K > 0$ the rigidity, $\tau > 0$ the pre-strain in the virtual elastic wire and $g_v$ the maximal gap above which the interaction vanished. The rigidity $K$ was defined as a force $F$ per strain of elastic wire. Pre-strain $\tau$ introduced a non-zero tension between two nodes at the beginning of the simulation.

$$\begin{cases} g \in ](1-\tau)g_0; g_v[ \Rightarrow R = -K(\frac{g-g_0}{g_0}+\tau) < 0 \\ g \in [0; (1-\tau)g_0] \text{ or } g \geq g_v \Rightarrow R = 0 \end{cases} \quad (1)$$

Contact law, which is based on classical Signorini and Lennard-Jones models (Jean, 1999), introduced frictionless contact between rigid spherical envelopes surrounding the nodes. To ensure the impenetrability of the envelopes of two interacting nodes, Contact law generated between nodes a compression force $R$ which was high enough to maintain them at a distance which could not be lower than the sum of the radiuses of their envelopes, $g_i$ (Eq. 2). Contact law was cohesive: when the envelope of two interacting nodes separated, a constant tension $c$ occurred, maintaining the interaction until a maximal gap $g_v$ above which it was broken.

$$\begin{cases} g = g_i \Rightarrow R > 0 \\ g \in ]g_i; g_v] \Rightarrow R = -c \\ g > g_v \Rightarrow R = 0 \end{cases} \quad (2)$$

Soft Contact law, an alternative form of the Contact law given by the set of Eq. 3, introduced softer compressive interactions between neighbouring nodes, making it possible for their virtual rigid envelopes to interpenetrate. As the node envelopes interpenetrate, the compressive force $R > 0$ increases linearly with the penetration ($g_i$ - $g$) according to a compressive rigidity $S > 0$.

$$\begin{cases} g \leq g_i \Rightarrow R = S.(g_i - g) \\ g \in ]g_i; g_v] \Rightarrow R = -c \\ g > g_v \Rightarrow R = 0 \end{cases} \quad (3)$$

To reproduce in the CDM model the various components of the CSK and NSK, specific interaction laws derived from the three basic laws, Elastic Wire, Contact and Soft Contact, were introduced between the various species of nodes. These specific interaction laws were defined with different values of the parameters ($S$, $K$, $\tau$, $g_v$).

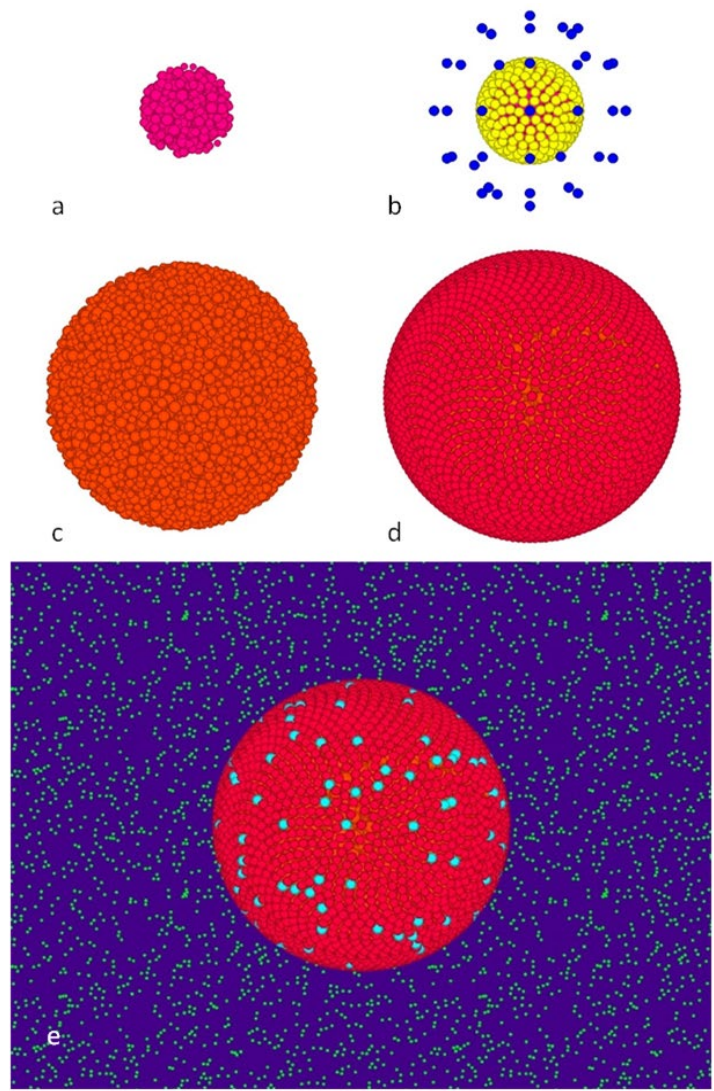

**Fig. 1**. Structure of the CDM model at round state (diameter = 15 μm). Cell geometry was represented by a divided medium whose nodes were composed of different species: (**a**) nuclear core (pink), (**b**) nuclear membrane (yellow) and perinuclear nodes (blue) (**c**) cell core nodes (orange) (**d**) cell membrane (red). (**e**) CDM model implemented into free adhesion simulation. The CDM was equipped with 150 adhesion complexes (cyan) in the cell membrane (red) and adhered on a flat substrate coating with extracellular adhesion proteins (green).

The 5 following laws of interaction were derived from Elastic Wire law. High-tensile LAMIN interactions were introduced between the nodes in the nuclear cortex to reproduce in the CDM model the nuclear lamina of the NSK. Intra-nuclear filament tensile interactions, so-called INTRA-NSK, were also introduced between nuclear core nodes to represent the intra-nuclear filaments made of lamin and actin. The actin filaments in the CSK were represented in the CDM model by low-tensile interactions, so-called short F-ACTIN, which created a cortical network between the nodes in the cell cortex as well as a diffuse network between cell core nodes. Actin filaments that form stress fibres were represented in the model by high-tensile interactions, so-called long F-ACTIN, connecting adhesion complexes to perinuclear nodes and cortex nodes. One stress fibre was thus represented by a bundle of several long F-ACTIN interactions. For instance, long F-ACTIN was defined by a rigidity K of $1.10^{-2}$ nN/ strain and a pre-strain $\tau$ fixed at 0.2 following the level of retraction observed in isolated stress fibres by Deguchi *et al.* (Deguchi *et al.*, 2006). Intermediate filaments form a dense and unstretched network around the nucleus and connect it to the actin CSK (Green *et al.*, 1986). They were represented in the model by INTER.FIL tensile interactions which connected the long F-ACTIN network to the nuclear LAMIN network and which were slackened in the initial state with a pre-strain $\tau$ of -0.1. Microtubules, which are known to resist and bear compression forces in cells (Kurachi *et al.*, 1995; Brangwynne *et al.*, 2006), were represented by MICROTUBULE law which was derived from Soft Contact law and generated compressive force networks in the whole medium of the CDM model. The CONTACT law, which was also derived from Contact law, introduced non-penetrability between the envelopes of nuclear nodes and governed the contact between the cell model and the substrate.

The tensile interactions that appeared in the CDM model during simulation represent the mechanical forces that are generated and supported in cells by actin filaments and intermediate filaments. Compressive interactions, which were computed as preventing interpenetration of virtual rigid envelops surrounding nodes, represent the compressive forces supported principally by microtubules, but also by intermediate filaments and the cytosol. So as to mimic CSK reorganisation during cell adhesion, the networks of interactions in the CDM model had variable connectivity. For instance, compressive interactions that were defined between close nodes could appear and disappear easily during cell deformation and created a diffuse network as represented by the labile microtubules. On the contrary, stress fibres and intermediated filaments, which are more stable, were represented in the model by networks of interaction that appeared between remote nodes and were less affected by disassembling in global deformation.

The mechanical behaviour of the CDM model was characterised by both the equations for the interaction between the constitutive nodes and the dynamic equation for the whole set of nodes (Milan *et al.*, 2007). The interactions between nodes depended on the gaps which separated them from each other and which could be calculated from the velocities and initial positions of all the nodes in the medium. The CDM model equations were solved using the LMGC90 code, which provides an open platform for modelling contact or multi-physics interaction problems. The LMGC90 code is based on Non-Smooth Contact Dynamics (NSCD), a numerical method which was developed for dealing with connecting elements in a divided medium (Dubois and Jean, 2006). The discrete forms of the dynamic equation of the whole set of nodes were solved along with interacting laws using an implicit non-linear Gauss-Seidel algorithm (Jean, 1999; Milan *et al.*, 2007) until the equilibrated state was reached.

In the present model, the number of interactions was neither fixed nor limited a priori but instead resulted indirectly from the value of gv, the maximal gap of interaction validity: an increase or a decrease in $g_v$ in a given law will respectively increase or decrease the number of interactions.

*Computation of intracellular tonus in in vitro cells depending on their adhesion conditions*
For each of 4 hMSC cell specimens of typical morphologies, FAPs were immunostained and their positions were measured and reported in the model. The cell model that was originally round, as displayed in Fig. 1a-d, entered into contact with a plane substrate and spread until the FAPs coincided with those observed experimentally. To do so, the nodes of the membrane that were closest to the final focal adhesion sites were chosen as candidates of trans-membranal adhesion complexes. The candidates were displaced iteratively toward the final location of the focal adhesions. The spreading process was performed over 50 iterations, the candidates covering at each iteration 1/50 of the distance between their initial and final positions. At each iteration, trans-membranal adhesion candidates were constrained at their current position and connected to each other and to the perinuclear nodes following long F-ACTIN laws mimicking formation of stress fibres. The number of long F-ACTIN interactions was nevertheless not limited. To compute valid intracellular tonus, the long F-ACTIN law rigidity, $K = 1.10^{-2}$ nN/strain, was defined a posteriori so that the tensile force supported by each single adhesion complex node remained within a range of 10-30 nN as reported by Balaban *et al.* (Balaban *et al.*, 2001).

To follow the evolution of internal cell tension during spreading, we therefore introduced an index known as intracellular tonus *T* which is computed as the sum of all Elastic wire interaction forces through the plane located in the middle of the cell and perpendicular to the direction of maximal intracellular tension. In the same way, intranuclear tonus TNSK was also defined to compute the tensile mechanical behaviour of the nucleus during cell spreading.

*Indexes of nuclear deformation*
The NSK was connected to the CSK and was able to deform as a result of CSK reorganisation (Itano *et al.*, 2003; Wang *et al.*, 2009). Nuclear deformation of adherent cells was measured experimentally as the ratio between horizontal major (M) and minor (m) axes. The nuclear deformation in the cell specimens of the typical morphologies was compared to that obtained in the cell model. As the cell model is three-dimensional, it gave access to the z-axis deformation of the nucleus. To describe the three-dimensional deformation of the nucleus, the octahedral shear strain $\varepsilon_{shear}$ was computed during adhesion of the cell model as a norm of differences between strains of the nucleus in the 3 directions of the space (Eq 4):

$$\varepsilon_{shear} = \sqrt{(\varepsilon_x - \varepsilon_y)^2 + (\varepsilon_x - \varepsilon_z)^2 + (\varepsilon_y - \varepsilon_z)^2} \quad (4)$$

*Free adhesion simulation*
In this part, the cell model, which was initially round, was able to adhere and spread on the substrate by detecting and freely connecting the adhesion proteins distributed randomly on the substrate surface (Fig. 1e). The trans-membranal adhesion complex nodes were able to bind every extracellular protein located at less than 1 μm by ADHESION cohesive contact law (c = 1nN). When an adhesion complex node became activated, it formed a FAP interacting with others *via* long F-ACTIN.

To mimic the process of cell adhesion and spreading by filopodium emission, an algorithm was introduced so as to modify iteratively the model's geometry. At each iteration, filopodia that were represented by moving nodes in the membrane were emitted on to the substrate surface at 1.5 μm away from existing FAPs. The mechanical state of the model was then computed until its equilibrated state, taking into consideration that the moving nodes could interact with extracellular proteins by the ADHESION law and with other neighbouring FAPs by the long F-ACTIN law. If the moving node, which represented the filopodium extremity, bound proteins, it formed a new FAP on a more distant adhesion site, and the initial FAP was deactivated. If the moving node did not bind any proteins in the substrate, it reintegrated the cell membrane or joined neighbouring focal nodes due to long F-ACTIN repel.

Table 1. Area, length, number of FAPs per surface and nuclear strains measured experimentally for each type of cell shape.

| Cell shape | Area in µm² | Length in µm | Number of focal points per cell | Nuclear strain (M/m) |
|---|---|---|---|---|
| Round | 803 ± 70 | 36 ± 3 | 21 ± 2 | 1.45 ± 0.2 |
| Branched | 1267 ± 50 | 109 ± 1 | 64 ± 1 | 1.36 ± 0.2 |
| Elongated | 832 ± 90 | 99 ± 1 | 52 ± 1 | 1.54 ± 0.2 |

Table 2. Experimental measurements and computational estimations of morphological and mechanical properties in the 4 cell specimens of round, spread, elongated and star-like shapes as well as of the CDM model implemented in the free adhesion process.

| Cell morphologies | Round | | Spread | | Elongated | | Star-like | | Free adhesion simulation | |
|---|---|---|---|---|---|---|---|---|---|---|
| | *In vitro* | Comput. | *In vitro* | Comput. | *In vitro* | Comput. | *In vitro* | Comput. | 20th it. | 50th it. |
| Cell diameter (µm) | 37.7 | 37.7 | 51.12 | 51.12 | 140 | 140 | 181 | 181 | 63 | 144.4 |
| Spread area (µm²) | 846 | 796 | 1,365 | 1,387 | 3,516 | 3,694 | 8,388 | 8,556 | 1,916 | 12,340 |
| Number of FAPs | 36 | 36 | 121 | 121 | 250 | 250 | 143 | 143 | 77 | 51 |
| Cell strain ($\Delta d/d_1$) | | 1.51 | | 2.41 | | 8.33 | | 11.07 | 3.2 | 8.63 |
| Focal tension (nN) | | -3.6 | | -3.4 | | -7.2 | | -15.2 | -4.0 | -21.3 |
| Long F-ACTINs / PFA | | 133 | | 116 | | 79 | | 103 | 164 | 257 |
| Intracell. tonus (nN) | | 24 | | 58.4 | | 192 | | 254 | 133 | 548 |
| Nuclear strain (M/m) | 1.5 | 1.00 | 1.20 | 1.10 | 1.59 | 1.88 | 1.33 | 1.35 | 1.09 | 1.09 |
| Nuclear strain ($\varepsilon_{shear}$) | | 0.45 | | 0.60 | | 0.95 | | 1.07 | 0.52 | 1.04 |

The spreading process was then able to continue until it reached, for instance, the final intracellular tonus previously found and corresponding to a specific morphology. The morphology of the cell model at the end of the adhesion process was then discussed and compared to the experimental observations.

**Results**

**Image analysis of cell morphology**
The adherent hMSC cells were imaged after 4 h of culture on TCPS. For the three different cell shapes: round, branched and elongated, which were considered, the results of area, length, number of FAPs and nuclear strain are summarised in Table 1. For each cell shape category, one specimen of adherent cell was chosen; in the case of the branched shape category, two specimens were chosen: one spread and the other star-like shaped (Fig. 2).

**Intracellular tonus depending on adherent morphology**
The CDM model was used to compute the mechanical state of cell specimens (Fig. 2). Results are presented and compared in Table 2. It is to notice that in star-like configuration, deformation of the model, reported as diameter variation, reached more than 1,000 %.

During spreading, the mechanical state of the model changed and involved an increase of active interactions. At initial state 50,000 interactions delivered non-null force while for instance, in the star-like morphology, they are 88,000. Considering the 4 morphologies, the more spread the model was, the more short and long F-ACTINs and INTER.FILs there were and the less MICROTUBULEs there were (Table 3). The magnitude of these tensile and compressive interactions ranged between 1-750 pN, highest magnitudes being found in the star-like morphology. Focal adhesion tensions averaged 10 nN (Table 2). It appears that remote FAPs withstood greater tensile force (Fig. 3). While most FAPs pulled the cell toward the periphery, some closed to the centre, withstanding tension forces from peripheral focal nodes, and exerted tension in the direction of the cell centre.

During adhesion and spreading of the model reaching the 4 cell morphologies, the creation of new long F-ACTINs connecting focal adhesion sites, as well as the stretching of all tensile interactions, led to an increase in intracellular tonus (Table 2 and Fig. 4a). With an initial intracellular tonus of 3.5 nN, the cell model reached tonus that was 6, 17, 55 and 73 times higher in round, spread, elongated and star-like morphologies, respectively.

NSK and CSK were linked in the model by INTER.FIL interactions. During cell spreading the nucleus deformed as did the CSK. Being defined as the ratio between major

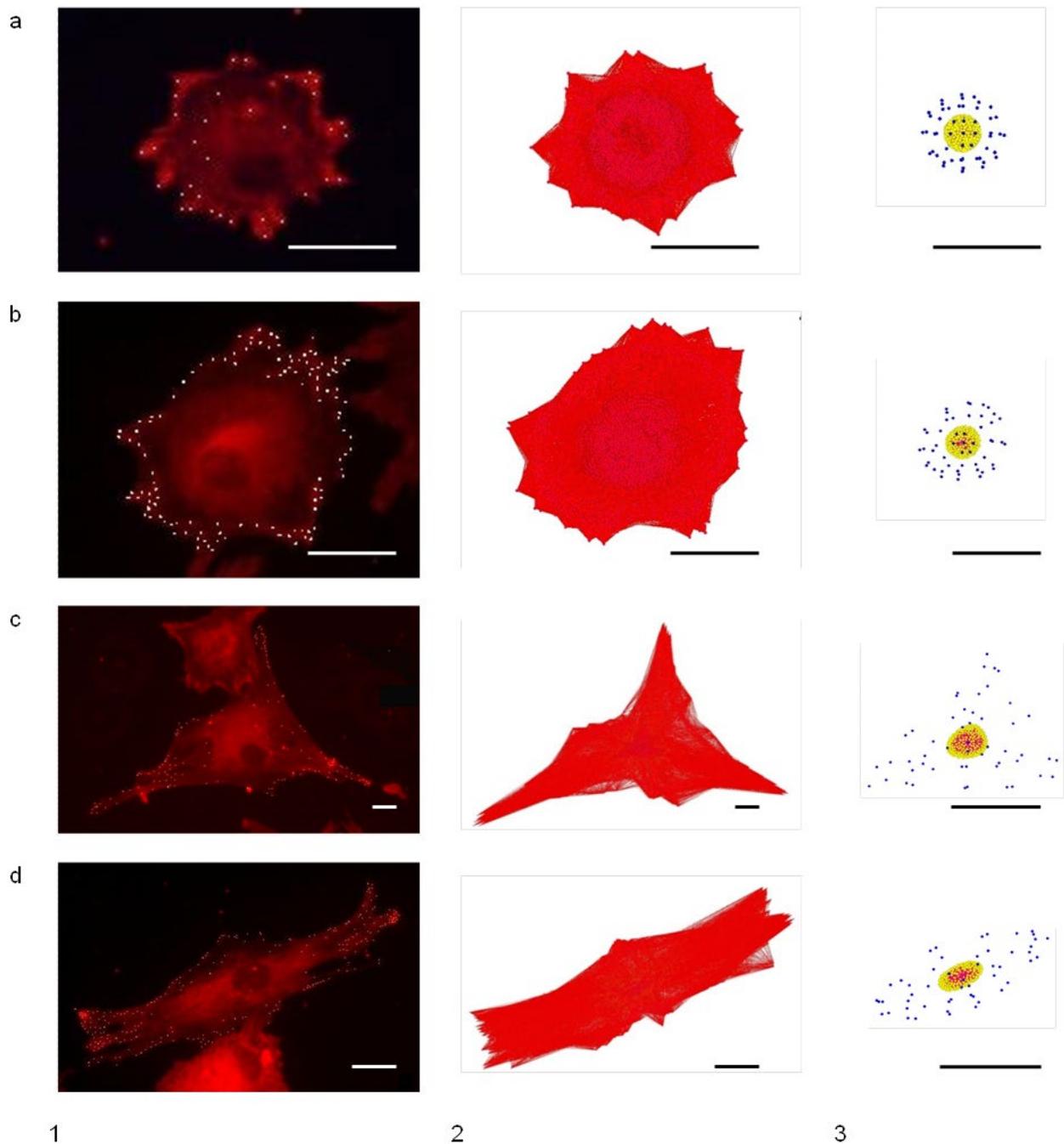

1  2  3

**Fig. 2**. In column 1 are shown the cell specimens of round (**a**), spread (**b**), star-like (**c**), and elongated (**d**) morphologies (bars = 20 μm). Focal adhesion points are shown as white points. In column 2 and for every morphological configuration, are shown in red the network of all tensile interactions within the spread state of the cell model that connects the substrate with the same adhesion conditions as the cell. In column 3 is shown in yellow (membrane) and red (core) the nucleus in the cell model surrounded by perinuclear nodes (blue).

and minor horizontal axes (M/m), the nuclear deformation computed in the model was similar to the one measured *in vitro*, especially in star-like morphology (Table 2). The model predicted the highest nuclear deformation in elongated morphology, which was observed *in vitro*. Round cell, which possibly initiated division, had a nucleus that was more deformed than that in the model. Except for round cell, *in vitro* and computational results showed that nucleus was more deformed in more spread morphologies. This was confirmed by the computation of $\varepsilon_{shear}$, the 3D deformation of the nucleus. Contrary to M/m, $\varepsilon_{shear}$ indicated that nuclear deformation was slightly lower in elongated morphology than that in star-like one. Fig. 4b shows, for the 4 cell morphologies, the evolution of nuclear deformation computed as shear strain during iterative spreading of the CDM model. Because the nucleus was surrounding by INTER.FILs that was initially unstrained and which became stretched after the cell model reached a certain level of deformation, it began to deform after a phase shift of roughly 5 iterations of the spreading process.

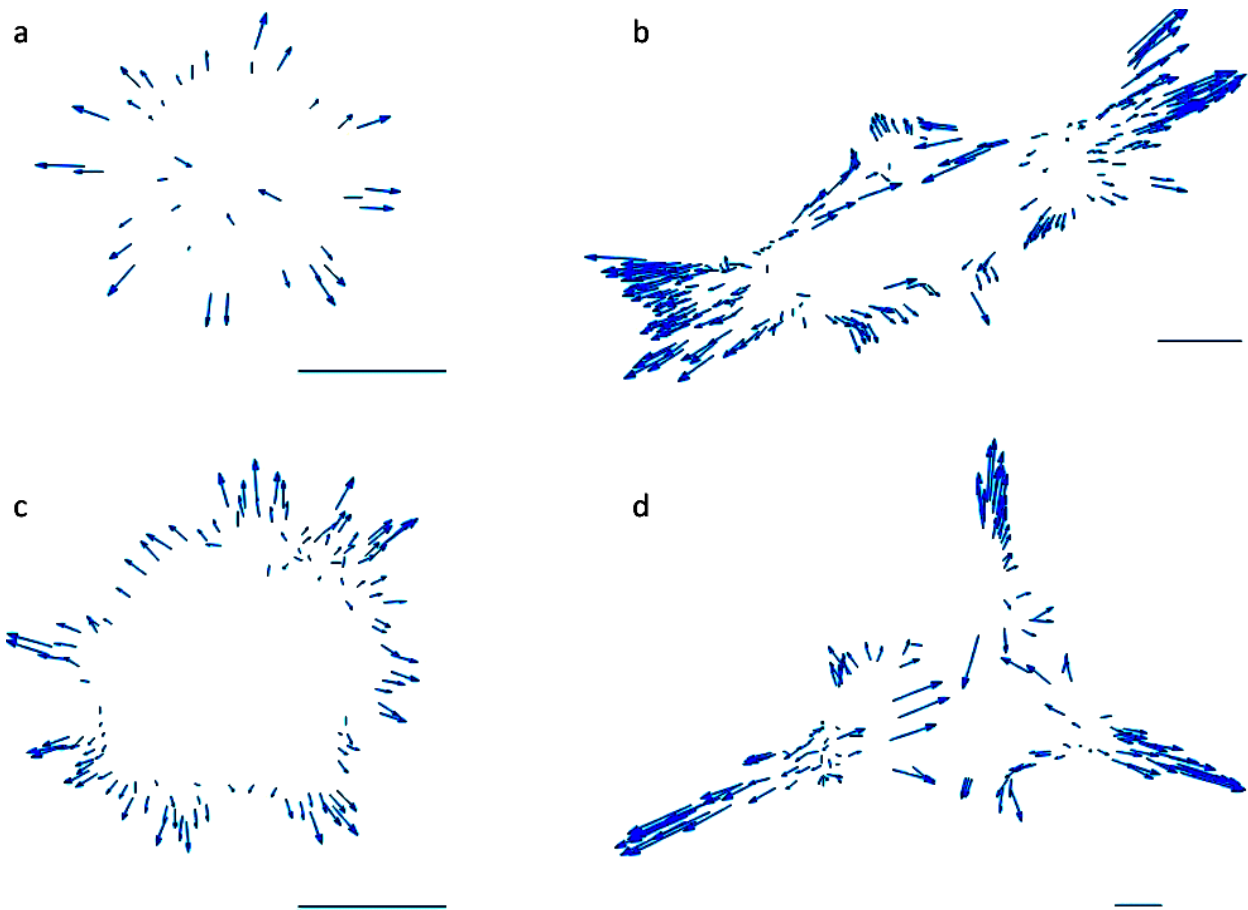

**Fig. 3**. Fields of focal adhesion forces, computed using the CDM model, for the 4 cell specimens, (**a**) round, (**b**) elongated, (**c**) spread, (**d**) star-like. Vectors represent tensile force exerted on the CSK by focal adhesion complexes. Bars = 20 μm and 10 nN.

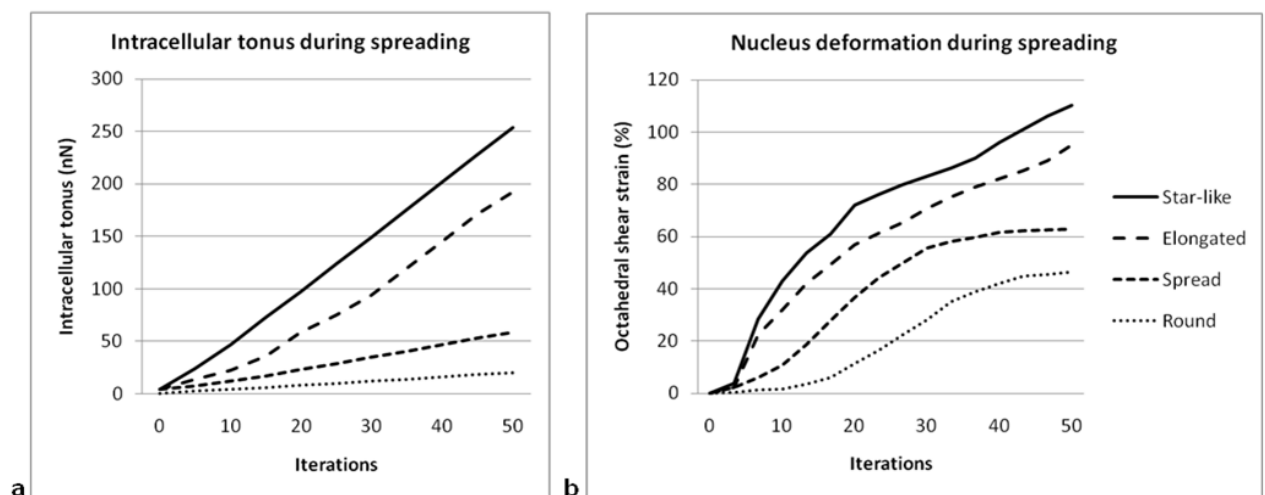

**Fig. 4**. Evolution of (**a**) intracellular tonus and (**b**) octahedral shear strain of the nucleus during the iterative adhesion process of the CDM model reaching the same adhesion conditions as the 4 cell specimens.

Then it deformed as intracellular tonus increased. Intra-nuclear tonus reached 0.2, 0.56, 2.8 and 3 nN for round, spread, elongated and star-like morphologies.

**Free adhesion simulation**
The CDM model spread here on a flat substrate coated with adhesion proteins, being driven by the algorithm of filopodium emission and focal adhesion creation (Video 1 – available on the paper's eCM Journal web page). As reported in Table 2 and shown in Fig. 5, the model underwent considerable deformation during the spreading process without CSK disrupting. Filopodium emission procedure may not result in substrate protein binding. If it did not, the filopodium may be redirected to neighbouring active adhesion sites and may create a new focal node within an existing adhesion site. This led to the formation of focal adhesion plaques composed of several focal adhesion nodes. As the model spread, the adhesion plaques were less numerous but stronger, being composed of more focal adhesion nodes and more ADHESION interactions with substrate proteins (Table 2 and Fig. 6). During spreading simulation, CSK reorganised with an increasing number of all interactions but the MICROTUBULEs that decreased dramatically in number from the 30th iteration as the model reached a large spread shape (Table 3). Compared with its initial value, intracellular tonus increased by 40 after 20 iterations of free adhesion simulation and by 160 after 50 iterations (Table 2). Some intracellular tonus was transmitted to the nucleus and induced its deformation. At the 50th iteration of the adhesion process, the octahedral shear strain of the nucleus, as well as the ratio between major *vs*. minor axes, was more than 100 % (Table 2). Intracellular tonus and nuclear strain increased in a non-linear manner with cell spreading, with nonetheless quasi-linear behaviour after the 30[th] iteration (Fig. 7). As described under the subheading "Intracellular tonus..", the spreading of the model led to nuclear deformation after decay caused by the initial slackened state of INTER.FILs which surround and connect the nucleus. In addition, as the cell model became thinner, the nucleus flattened and was pushed toward the substrate. Deformation of the nucleus induced the apparition of intranuclear tonus equal to 5 nN at the 50[th] iteration.

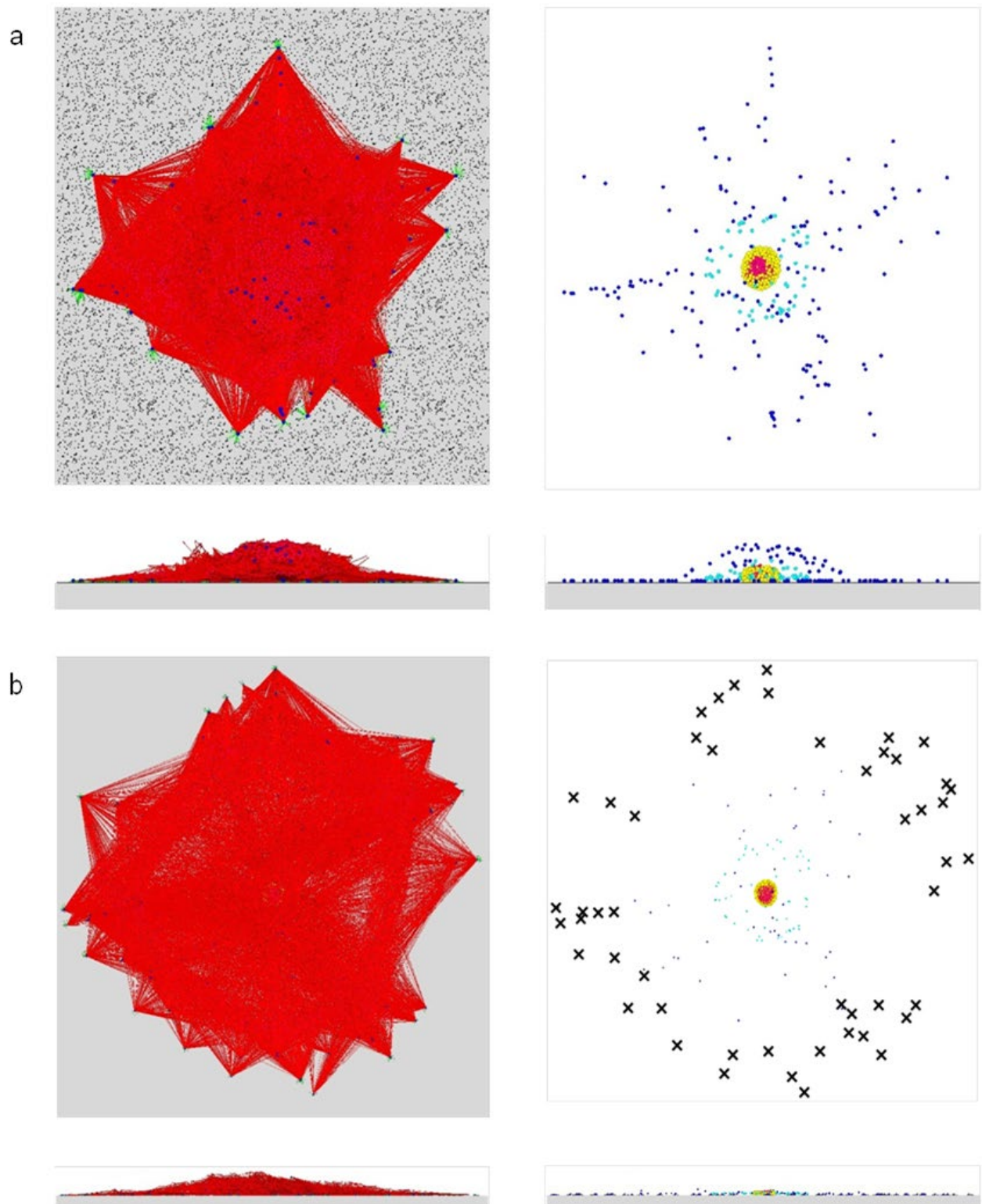

**Fig. 5**. The cell model during free adhesion simulation, conducted by a filopodium creation algorithm on a flat substrate coated with extracellular adhesion proteins. (**a**) Spread state of the cell model after 20 iterations; cell diameter = 63 μm. (**b**) Spread state after 50 iterations; crosses indicate adhesion plaques, cell diameter = 144.4 μm.

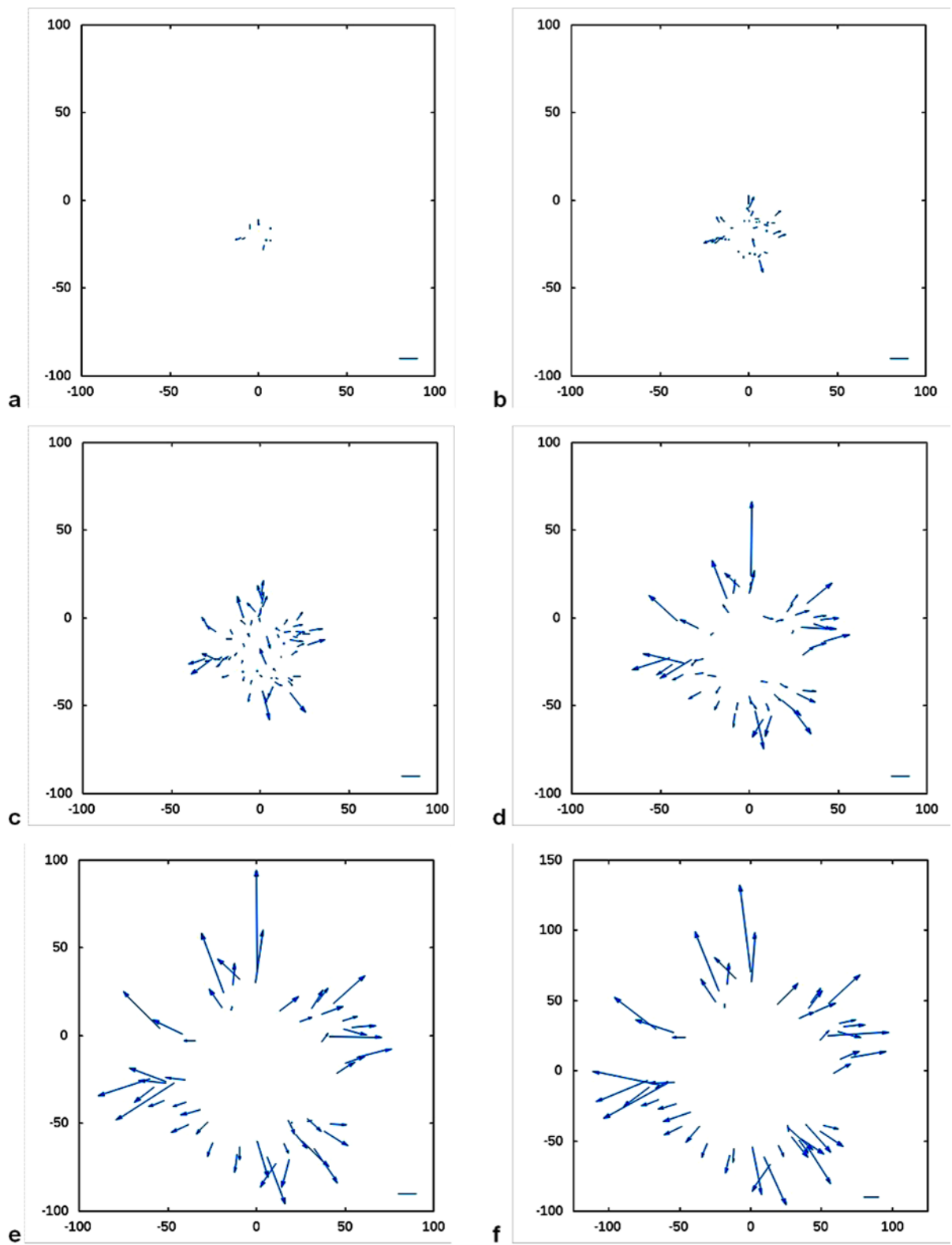

**Fig. 6**. Field of adhesion plaque force vectors, over the iterative cell model free spreading process, at iterations 2 (**a**), 10 (**b**), 20 (**c**), 30 (**d**), 40 (**e**) and 50 (**f**). Bar = 10 nN. Scales are given in μm. The number of adhesion plaques, the average adhesion force and the maximum force are given as follows. Iteration 10: nbr 38, mean 1.6 nN, max 7 nN; iteration 20: nbr 77, mean 4 nN, max 16.4 nN; iteration 30: nbr 57, mean 10.1 nN, max 42.2 nN; iteration 40: nbr 53, mean 15.4 nN, max 56.3 nN; iteration 50: nbr 51, mean 21.3 nN, max 62 nN.

The cell model at the 20[th] iteration of free adhesion simulation was deformed the same as it did in spread morphology (Results, subheading "Intracellular tonus..", Table 2). Nuclear strain was also the same while intracellular tonus was higher. At the 50[th] iteration, the cell model had a spread area larger than that in the star-like morphology, with a more circular shape and a smaller diameter. It attained intracellular tonus that was 2 times higher with the same nuclear octahedral strain.

### Discussion

This study proposed a combination of experimental observations and a computational model as a way to identify the influence of adhesion conditions on cell morphology, CSK tension and finally on the mechanical behaviour of the cell. Adhesion dynamics induce complex regulation of biochemical reactions such as integrin binding to the extracellular matrix. Extracellular or intracellular physical pulling forces on focal adhesions induce tyrosine-phosphorylation of the GPTase Rho initiating their growth and strengthening by the recruitment of proteins (Riveline *et al.*, 2001; McNamara *et al.*, 2010), which illustrates the mechanosensitivity of the focal adhesion sites. In our study, we analysed the focal adhesions of hMSCs on TCPS by staining vinculin, which links integrin to the actin CSK. It has been shown that tensile force could also modify the shape of talin, allowing it to bind additional vinculin (Del Rio *et al.*, 2009). Moreover, integrin mediates signalling pathways between cells and the matrix, activating both direct mechanotransductive signalling *via* the CSK and an indirect molecular cascade that regulates transcription factor activity, and gene and protein expression. The CSK Divided-Medium (CDM) model that is proposed here computes the mechanical forces transmitted by the CSK from the focal adhesions to the deformable nucleus and would thus represent the direct mechanotransduction involved in cell adhesion.

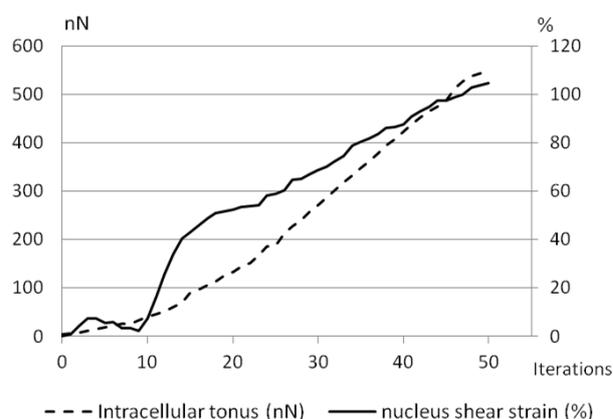

**Fig. 7**. Evolution of intracellular tonus and octahedral shear strain of the nucleus computed in the cell model over 50 iterations in the free adhesion process.

**Video 1**. (available on the paper's eCM Journal web page) Iterative simulation of autonomous adhesion of the CDM model on a coated substrate. The tensile CSK is shown in red. The CDM, which is initially round (diameter =15 μm), is able to emit filopodia so as to detect and connect to the extracellular adhesion proteins distributed randomly on the flat substrate. After 50 iterations it reaches a highly spread morphology of diameter = 144 μm.

**Table 3**. Number, mean force and force max of the interactions which appeared between the nodes making up the CDM model and which represent the mechanical action of CSK filaments.

|  |  |  | 4 morphologies |  |  |  | Free adhesion |  |
|---|---|---|---|---|---|---|---|---|
|  |  | Initial | Round | Spread | Elongated | Star-like | Iter. 20 | Iter. 50 |
| **Number** | MICROTUBULE | 3,629 | 20,753 | 12,273 | 5,760 | 4,618 | 10,676 | 1,412 |
|  | Short F-ACTIN | 35,889 | 35,555 | 38,672 | 42,586 | 45,911 | 43,230 | 47,799 |
|  | Long F-ACTIN | 0 | 4,695 | 13,588 | 19,513 | 14,656 | 12,663 | 13,119 |
|  | INTER. FIL | 6,914 | 12,850 | 15,154 | 17,892 | 18,293 | 18,401 | 21,027 |
|  | LAMIN | 1,051 | 961 | 1,047 | 1,160 | 1,276 | 1,067 | 1,290 |
|  | INTRA-NSK | 2,349 | 1,164 | 1,577 | 2,006 | 2,727 | 1,691 | 2,767 |
| **Mean force (pN)** | MICROTUBULE | 1.1 | 1.2 | 2.0 | 1.8 | 4.9 | 4.8 | 8.5 |
|  | Short F-ACTIN | 0.0 | 0.2 | 0.3 | 0.7 | 1.1 | 0.5 | 1.4 |
|  | Long F-ACTIN | 0.0 | 13.9 | 15.2 | 48.6 | 73.8 | 30.4 | 96.9 |
|  | INTER. FIL | 0.2 | 4.7 | 9.8 | 30.3 | 41.7 | 12.9 | 56.3 |
|  | LAMIN | 0.7 | 2.4 | 3.8 | 10.0 | 10.1 | 5.1 | 10.9 |
|  | INTRA-NSK | 0.3 | 0.7 | 1.5 | 6.0 | 6.2 | 2.1 | 5.6 |
| **Force max (pN)** | MICROTUBULE | 10.7 | 47.8 | 128.1 | 79.4 | 146.5 | 636.8 | 150.7 |
|  | Short F-ACTIN | 0.3 | 1.9 | 3.2 | 18.8 | 22.9 | 6.6 | 17.4 |
|  | Long F-ACTIN | 0.0 | 68.1 | 89.5 | 478.8 | 728.9 | 209.9 | 586.3 |
|  | INTER. FIL | 3.1 | 43.6 | 74.8 | 292.0 | 454.1 | 104.6 | 311.6 |
|  | LAMIN | 9.2 | 22.5 | 28.8 | 41.2 | 44.4 | 34.5 | 45.1 |
|  | INTRA-NSK | 1.6 | 5.0 | 8.7 | 22.0 | 22.8 | 11.6 | 21.8 |

**Description of CSK components by the CDM model**
During simulation of the adhesion process, the tensile network evolved with the appearance of long F-ACTINs between new focal adhesion nodes, each focal adhesion node being involved in 100-200 long F-ACTIN interactions. In cells, these numerous long F-ACTINs may represent bundles of actin filaments which gather together and form one or more stress fibres per focal adhesion. Long F-ACTINs were prestrained in the 3D model at 20 % consistently with the retraction of isolated stress fibre (Deguchi *et al.*, 2006).

Intermediate filaments are known to resist major deformations, whereas they are less involved in small cell deformations (Wang and Stamenović, 2000; Fudge *et al.*, 2003). INTER.FILs, which represent intermediated filaments, were given a negative pre-strain of -10 % as in former CSK models (Wang and Stamenović, 2000; Sultan *et al.*, 2004). This means that the interactions were null around the nucleus at the non-deformed state and became activated following significant deformation of the cell model.

The numerous tensile and compressive interactions, which appeared in the CDM model, represented the mechanical action of the filaments of the CSK; one interaction corresponded to one filament. For instance, considering the modelling of actin CSK, during computation, roughly 45,000 1 µm-length short F-ACTINs and 15,000 20 µm-length long F-ACTINs were established in the model (Table 3). Bearing in mind that 370 monomers of actin are needed to form one 1 µm-length microfilament (Boal, 2002), the CDM model would contain 130,000,000 molecules of polymerised actin (F-actin). These amounts of F-actin are consistent with those measured experimentally: a cell can contain 500,000,000 actin molecules, half being F-actin (Lodish *et al.*, 2000).

The mechanical state of the model resulted on the one hand from the distribution of focal adhesion, as observed in the *in vitro* cells, and, on the other, from the magnitude of focal adhesion tensile forces that was defined according to the literature. As a result, the forces of the various interactions in the CDM model may consistently represent the level of internal forces born by CSK filaments individually. For instance, when the CDM model was highly spread such as in the star-like morphology, F-ACTINs, long F-ACTINs, INTER.FILs and MICROTUBULEs bore mean forces equal to 1.1, 74, 42 and 4.9 pN and maximal forces equal to 23, 730, 454 and 147 pN, respectively (Table 3). Freitas reported failure strengths of ~170 pN for actin microfilaments, ~20,000 pN for intermediate filaments and buckling failure of a 1 µm-length microtubule for a compressive force of 340 pN (Freitas, 1999). Although certain maximum forces may be in the same range as the failure strengths of the filaments, the mean forces were far lower and can thus be considered as valid.

Many other approaches, such as the finite element (FE) method, were proposed to model cell mechanics. FE models that are based on continuum field theory took well into account the cytosol and the cell membrane but they hardly described the highly-structured CSK in adherent cells. Nonetheless, homogenised theories of pro-elasticity, soft glass materials or gels could be used in FE models so as to compute overall mechanical behaviour of cells with local heterogeneous description of cytoplasm. FE models could also be built from 3D reconstruction of real CSK structure imagined by confocal microscopy; CSK filaments or fibres would be modelled one by one as 3D rode structures. Besides the high numerical cost, this approach would not allow CSK rearrangement as easy as in CDM model by forming or vanishing filaments. In fact, numerous FE cell models started to integrate identical descriptions of filaments, other than that in CDM model, in using connector elements that link two nodes to each other following elastic interaction law (Slomka and Gefen, 2010).

Compared to the continuum approach of finite element (FE) modelling, the advantage of the CDM model is to describe easily, by means of the LMGC90 code, the discrete filamentous structure of CSK. The mechanical interactions in the CDM model represent the localisation and amount of CSK filaments. For instance, CSK maturation at the end of adhesion can be described by an increasing number of stress fibres between FAPs in controlling only interaction length. The second advantage of the CDM model is the description of CSK rearrangement in including variable mechanical connectivity. Nonetheless, if the present CDM model has overtaken some limits we exhibited in our previous study (Milan *et al.*, 2007) – such as 2D geometry and rigid nucleus, the fluid effect of cytosol needs to be modelled using multi-physics coupling inside the LMGC90 platform (Dubois and Jean, 2006). In addition, the mode of CSK rearrangement that is proposed here remains not clearly verified and should be validated.

**Mechanical state of adherent cell depending on its morphology**
McBeath *et al.* showed that changes in cell shape alone are enough to mediate the switch in hMSC commitment from their adipogenic and osteogenic fate caused by the ROCK-mediated cytoskeletal tension and may raise the possibility that contractile activity increased with cell spreading (McBeath *et al.*, 2004). We thus chose to investigate relations between intracellular tonus and cell shape. We confirmed that intracellular tonus, computed as the sum of tensile forces, increased with that of the spread area. For instance, the spread cell with a diameter of 51 µm and a spread area of 1,387 µm² had an intracellular tonus of 58 nN, while the star-like cell with a diameter of 181 µm and a spread area of 8,388 µm² had an intracellular tonus of 254 nN. The differences between mechanical states were caused not only by cell shape but also by the number of focal adhesions, which influenced the amount of long F-ACTINs in the model. Intracellular tonus, computed by the model, resulted from cell deformation and stretching as well as changes in CSK stiffness by the formation of stress fibres. Intracellular tonus as a function of adhesion conditions could explain the driving factor behind cell shape in cell differentiation (Kilian *et al.*, 2010; McBeath *et al.*, 2004).

The more the CDM model spread, the more the intracellular tonus was transmitted directly to the substrate *via* focal adhesions. So the part of tonus born by compressive MICROTUBULEs interactions decreases and they disassemble (Table 3). The reduced mechanical role of microtubules in spread cells was observed experimentally (Hu *et al.*, 2004). Besides, numerous studies showed that, in the other direction, disruption of MTs, using drugs, leads to cell spreading and creation of focal adhesions and stress fibres in involving the Rho signal cascade (Enomoto, 1996). Nonetheless, other studies showed that, during cell adhesion and spreading, the growth of microtubules targeted early FAPs to orient and stabilise them (Krylyshkina *et al.*, 2003). PFA-targeting microtubules in a biochemical purpose seems to have no mechanical action contrary to the hypothesis of Maurin *et al.* and this was not taken into account in the present CDM model (Maurin *et al.*, 2008).

In the present study, the differentiation of the adherent stem cells cultured *in vitro* was not analysed. The computational results referred to morphological data which are known to have an influence on cell differentiation. As a perspective, the CDM model could follow dynamically the adhesion process of a stem cell on micro-patterned substrate until differentiation, so as precisely to identify the intracellular mechanical cues and events inducing differentiation. In addition, the model could also predict CSK restructuring as a result of differentiation, by reinforcement and redistribution of stress fibres and focal adhesions.

Computing the mechanical state of adherent cells observed *in vitro* was based here only on the distribution of focal adhesions on the substrate surface. The focal adhesion forces were not measured and were introduced into the model as boundary conditions. Nonetheless, the resultant focal adhesion forces obtained in the model were in the same range as those published in other studies. For instance, they were equal to 15 nN, with peak of 60 nN, in the star-like morphology as well as at the 50$^{th}$ iteration of free adhesion simulation. In adherent fibroblasts, Balaban *et al.* measured focal adhesion forces equal on average to 10 nN with a peak of 30 nN, and identified a linear relation between force and area of focal adhesion of 5.5 nN/µm$^2$ (Balaban *et al.*, 2001).Tan *et al.* found an equivalent relation in smooth muscle cells  (Tan *et al.*, 2003). In hMSC cells, Fu *et al.* measured focal adhesion forces equal to 4-8 nN for areas of 1-2 µm$^2$ and on rigid substrates identified a relation force-area of 3.7 nN/µm$^2$ (Fu *et al.*, 2010). Using this last relation, the focal adhesions in the CDM model in the star-like morphology, or at the 50$^{th}$ iteration of free spreading, would have a mean area of 4 µm² (2.2 µm diameter).

**Free adhesion simulation**
In this experiment, the CDM model represented the motile activity of cells during the adhesion process by simulating filopodium emission and creation of FAPs connecting extracellular proteins on a flat substrate. In the model, creating focal adhesions induced the creation of pre-strained stress fibres that resulted in an increase in intracellular tonus and cell stiffness. Changes in CSK stiffness that is described here by the model may affect cell sensitivity to mechanical stimuli or substrate stiffness.

Focal adhesion forces were represented in the model by cohesive contact between focal nodes and substrate proteins, one focal node being able to bind every substrate protein located at less than 1 µm with a maximum tensile strength of a single bond of 1 nN. Because focal adhesion nodes interact with each other *via* high tensile force generated by stress fibres, if a new focal adhesion failed to connect strongly to the substrate, it was detached and joined neighbouring strong focal adhesions. This led to a gathering of focal nodes around strong points and the transformation of FAPs into adhesion plaques. Finally, although focal adhesions were represented in the model by simple cohesive contacts between integrins and proteins, the model predicted the spreading of adherent cells, taking into account the aggregation and remodelling of focal adhesions. A more realistic modelling of focal adhesion kinetics could be introduced into the model to predict the creation and maturation of adhesion complexes based on mechanochemistry (Yamada and Geiger, 1997).

In free adhesion simulation, some nodes in the membrane were emitted to 1.5 µm outside of the model and were able to connect with substrate proteins. These represented the filopodia, in a simplified way, with cytoskeletal processes mainly composed of actin filaments and transmembrane adhesion complexes. In the model, if filopodium nodes connected with extracellular proteins, they resulted in the creation of focal adhesions and the cell spread in that direction. Free adhesion simulation showed that the CDM model reached widely spread adherent shapes that were equivalent to those reported experimentally. It may thus consistently predict cell adhesion and make it possible to analyse the morphology of adherent cells, depending on substrate topography and surface protein distribution. This may help in distinguishing optimal surface treatments for biointegration of biomaterials. In future studies, an end of spreading criteria, such as intracellular tonus or nuclear strain, should be introduced in free adhesion simulation so that the cell model reaches viable morphologies. In addition, a process based on chemical and mechanical energy minimisation could be considered, so as to predict optimal morphology regarding the final energetic state as well as the history of energetic costs to reach such a state. In the present model, biological time was not taken into account. Only a mechanical time was considered after filopodium emission to compute CSK reorganisation and when mechanical equilibrium was reached. Nonetheless, spreading velocity in simulation can be deduced from the biological time of the adhesion process observed experimentally and ranges between 5-50 µm$^2$/min. In addition, future studies are needed to compute not only the mechanical equilibrated state of a cell under given adhesion conditions but also its mechanical response to external solicitations such as matrix strain or fluid flow.

**Nuclear deformation and mechanosensing**

Our results indicate quantitatively how the mechanical forces transmitted by the CSK to the NSK affect nuclear shape. They confirm the experimental results from other studies (Dahl *et al.*, 2008; Nathan *et al.*, 2011). Computations have shown here that nuclear strain increased in a non-linear manner with intracellular tonus and spread area. As stated by several authors, nuclear deformation may induce changes in gene expression and cell differentiation (Nathan *et al.*, 2011). Thus, by analysing relations between nuclear shape and the internal tension of the CSK, our study quantifies the first steps of indirect mechanotransduction involved in cell adhesion process. It has also identified specific levels of intracellular tonus and nuclear strain in typical cell shapes known to lead as a predominant factor to specific commitments (McBeath *et al.*, 2004).

In the CDM model, the horizontal area of the nucleus, 28 μm$^2$ at the initial state, increased by 1.06, 1.24, 1.14 and 1.9 times in round, spread, elongated and star-like morphologies, respectively, and increased 1.3 times at the 50$^{th}$ iteration of free adhesion simulation. Deformation in the z-direction cannot be neglected when analysing global nuclear strain. In the CDM model, the nuclear z-axis which was 6 μm at the initial state, decreased by 0.45, 0.57, 0.68 and 0.60 for round, spread, elongated and star-like morphologies, respectively, and by 0.49 at the 50$^{th}$ iteration of free adhesion simulation. These levels of nuclear deformation are consistent with the measurements reported by Itano *et al.* In their study, they observed that nuclei flattened as the cells adhered and spread (Itano *et al.*, 2003). The maximum nuclear deformation corresponded to an increase of 1.6 of the horizontal area and a decrease of 0.4 of its z-axis. They showed that those levels of nuclear deformation are related to variations in calcium concentration in the nucleus and specific gene transcription. Entry of calcium may be explained by the opening of transmembrane ion channels when the nucleus deforms. In the CDM model, variations in the gap between nodes of the nuclear membrane could be computed as nuclear membrane deformation. Nuclear membrane strained at 4.4, 4.7, 12 and 22 % in round, spread, elongated and star-like morphologies and 38 % at the 50$^{th}$ iteration of free adhesion simulation. These levels of membrane deformation computed in the CDM model are those sensed by ion channels, which would make it possible to quantify the effect of direct mechanotransduction on calcium entry into the nucleus. Besides, deformation of the nucleus induces repositioning of chromosomes and so affects gene transcription, as observed by McNamara in confining the nucleus by the CSK and substrate micro-grooved topography (McNamara *et al.*, 2012). Mechanosensors may be also introduced into the cell model at focal adhesion complexes or located on nodes of the CSK network. For instance, the mechanosensitive molecule talin, connecting integrin to the CSK, may be represented as a deformable unit initiating focal adhesion maturation and CSK stress fibre assembly.

**Conclusions**

The combination of the 3D CDM model with the *in vitro* culture of adherent stem cells has made it possible to quantify some mechanical changes that occur during cell adhesion and may constitute mechanotransduction signals in cell commitment. Considering that the morphology of adherent stem cells is one of predominant factors controlling their differentiation we identified, in stem cell specimens of typical shape, specific levels of intracellular tonus and nuclear strain which may be the mechanical cues for differentiation. The CDM model is also able to adhere to a substrate autonomously, such as a cell by reproducing the typical cellular mechanisms of filopodium emission and the creation of focal adhesions. The simulation led to adherent cell shapes, which are consistent with those reported experimentally and with equivalent mechanical state. The CDM model may thus be a useful tool for characterising cell adhesion on biomaterials, in relation to surface properties such as roughness of the micro/nano structure and coating. More generally, it may help the understanding of all cell activities involving cytoskeleton mechanobiology.

**Discussion with Reviewers**

**Reviewer I**: Did the authors verify the computed 3D cell shapes with optical measurements, such as confocal microscopy?
**Authors**: No, but it would certainly add much to the study to compare e.g., the height of the nucleus in adherent cells and in the model. We cited, however, the work of Itano *et al.* (2003, text reference) that measured nuclear height using confocal microscopy. They obtained the same level of nuclear strain for equivalent cell spreading.

**Reviewer I**: If intracellular tonus is dependent on adhesion conditions, would changing the proteins adsorbed onto the TCP substrate alter the results of this study? The mechanical properties of cells have been shown to be highly dependent on the substrate (not just compliancy, but also ligand types). This could be a confounding factor in the study since serum proteins were allowed to freely adsorb during experimentation.
**Authors**: In this study, we computed intracellular tonus related only to the distribution of focal adhesion points. Different ligand types, ligand densities and/or substrate topographies would lead the cells to adhere with possibly less or more focal adhesion points. Besides, this may affect cell morphology and can result in different proportions of cells being placed into classified morphologies. However, in such a case, we would have computed intracellular tonus in the same way, as related only to the distribution of focal adhesion points. In such a case the same methods can be used.

Nonetheless, the reviewer is right, ligand type does not only influence the number of focal adhesion points. It also has influence on the strength-quality of the integrin-ligand bond. Locally, at the nm scale, a bond between one integrin and one ligand may have a different tensile strength-quality, depending on the integrin type and ligand type. In this study, to simplify, we considered a mean strength-quality of an integrin-ligand bond. We considered that the focal adhesion point, composed of several integrin-ligand bonds, has a strength that is related linearly to its area. In fact, in this study each focal adhesion point bears a tensile force ranging between 10-30 nN, as reported by Balaban *et al.* (2001, text reference). This range is commonly observed.

Controlling ligand type and density could be very interesting, but we think that computation of mechanical adherent cells first needs the experimental measurement of focal adhesion forces. This would allow computation with complete boundary condition (spatial distribution of focal adhesions plus real forces). The use of a flexible micropost substrate such as the one used by the group of Chen in Pennsylvania, may fit the objective. Using a micro-post substrate, together with a unique ligand such as fibronectin, will allow us completely to control adhesion. This will be the next step of our approach.

**Reviewer II**: What about shape of adhesion – e.g. in work by Biggs *et al.* (2007, additional reference) where he showed that adhesion length is important for osteogenesis. Is it important to bring such parameters into your model?
**Authors**: Elongation of focal adhesions is linked to their maturation and associate transductive pathways such as vinculin and focal adhesion kinase involved in osteoblastic differentiation. Using micro-nano-patterned surfaces, promoting focal adhesion elongation, may so influence osteoblastic differentiation. The point that the reviewer raises thus constitutes a key feature in the modelling of cell adhesion and its implication in osteoblastic commitment. We started to think how it could be described in the model. Currently, the model established a circular focal adhesion: one focal adhesion point interacts with all proteins located at less than 1 μm. We can see aggregation of several focal adhesion points to form an adhesion plaque. To represent maturation and elongation of focal adhesion points, we are considering adding nodes or particles, linearly, in a direction opposite to that of the intracellular force. The elongated adhesion would thus be represented by a chain of nodes. In considering maturation/elongation kinetics, we could attribute a different mechano-transductive behaviour to focal adhesions. We thus could simulate adhesion on biomaterial surface and quantitatively predict the size and shape of focal adhesions, to know if the topography would promote osteoblastic differentiation. In fact, we hope that this will be introduced in the next version of the model.

**Reviewer II**: What about inside the nucleus – i.e., as the nucleus changes shape what morphological / epigenetic changes might occur to control osteospecific differentiation?
**Authors**: As we noted in the paper, deformation of the nuclear membrane may lead to calcium entry and specific gene expression. In the same way, McNamara *et al.* (2012, text reference) have shown that nuclear strain may reorient chromosomes and promote specific transcription. This reorientation affects larger chromosomes that may contain osteoblastic genes. For now, the model represents the nucleoskeleton as composed of a highly reticulated network of lamin filaments in the lamina and a looser network of lamin and actin filaments on the inside. The looser network at the inside was not organised specifically. Neither chromosomes nor possible connection with nucleoskeletal filaments were taken into account inside the nucleus. A better description of the nucleus (nucleoskeleton, ion mechanosensitive channels, genetic material, etc.) could definitely improve the model. The mechanobiological processes involved inside the nucleus constitute the next subject of our research, combining experimental observations and numerical modelling.

**Reviewer II**: What about tensegrity – in what ways can this model be used to support this theory?
**Authors**: We discussed that point in our previous paper (Milan *et al.*, 2007, text reference), as well as the comparison with others theories. The CDM model results from equilibrium between tensile and compressive discrete interactions. Stiffness of the model increases with the internal tonus. In that way, the model follows the tensegrity concept. It is different from the first classical tensegrity model in allowing interaction reorganisation. This allows the modelling of complex processes such as adhesion in a biomimetic manner. Simulating adhesion using 3 or 6 bars tensegrity models would be limited: it would not be possible to introduce into the model the 100-200 focal adhesion points of a cultured cell as boundary conditions and it would not be possible to compute their mechanical adherent state. In fact, the CDM model is an alternative tensegrity model that is less limited than the classical ones. The CDM model brings the tensegrity concept further by including it inside a more biomimetic description of complex cellular process.

Nonetheless, the CDM model can be close to other theories such as the mechanotransduction theory of percolation (Forgacs, 1995, additional reference) as well as sol-gel theory, which describes reorganisation of CSK by transitional fluidifying before stabilising. Sol-gel theory is able to describe the migration process by fluidification and circulation of actin filaments. In the CDM model, fluidification can be generated by reducing the length of interaction between nodes and the corresponding gel state can be achieved by increasing the distance between interaction fixing nodes. In the CDM model, modifying dynamically and locally the interaction distance may describe cell motion and stabilisation.

**Reviewer II**: How do you predict using this model in biomaterials research?
**Authors**: To predict adherent cell behaviour usefully in biomaterials research, our approach would first need some validations and improvements. The mode of reorganisation of interactions that describe CSK filament reorganisation needs to be validated. The modelling of filopodia, lamellipodia and FA elongation need to be taken into account or improved. In addition, the model should integrate mechanosensors, and a relationship between coupling of mechanosensing level and biological response observed experimentally.

Nevertheless, currently, the CDM model can be used to study the influence of substrate topography on cell mechanical state. For instance, on 10-100 µm patterned substrates made of peaks and valleys, the model can compute precisely if the more tense cell shape is located on a peak, slope or in a valley. We can also analyse the influence of 0.1-5 µm patterned surfaces on focal adhesion creation and aggregation.

In addition, we can currently analyse the influence of adhesion density (size and spacing) on the cellular mechanical state (tonus, nuclear strain), as well as the mechanical effect of the design of coated surfaces (cross, triangle, circle or line).

The model can be used to compute the mechanical state of adherent cells in 3D polymeric matrices and to compare this with the 2D substrates. It may explain, perhaps due to a stiffer state, why cells are more mechanosensitive in 3D.

## Additional References